\documentclass[sigconf]{acmart}

\AtBeginDocument{%
  \providecommand\BibTeX{{%
    \normalfont B\kern-0.5em{\scshape i\kern-0.25em b}\kern-0.8em\TeX}}}

\copyrightyear{2022} 
\acmYear{2022} 
\setcopyright{rightsretained} 
\acmConference[FAccT '22]{2022 ACM Conference on Fairness, Accountability, and Transparency}{June 21--24, 2022}{Seoul, Republic of Korea}
\acmBooktitle{2022 ACM Conference on Fairness, Accountability, and Transparency (FAccT '22), June 21--24, 2022, Seoul, Republic of Korea}\acmDOI{10.1145/3531146.3533145}
\acmISBN{978-1-4503-9352-2/22/06}

\usepackage{amsmath}
\usepackage{bbm}
\usepackage{subcaption}
\usepackage{placeins} 
\usepackage{cleveref}

\newcommand\image{X_i}
\newcommand\hind{h_{ij}}
\newcommand\hagg{\overline{h_i}}
\newcommand\gt{y_i}
\newcommand\fhat{\hat f(X_i)}
\newcommand\fhatext{\hat f^{(\text{ext})}(X_i)}

\newcommand\gtcont{y_i^\text{(cont)}}

\newcommand\residual{r_i}

\begin{document}

\title{Trucks Don't Mean Trump:\\ Diagnosing Human Error in Image Analysis}


\author{J.D. Zamfirescu-Pereira}
\affiliation{%
  \institution{University of California, Berkeley}
  \city{Berkeley}
  \country{USA}}

\author{Jerry Chen}
\affiliation{%
  \institution{Stanford University}
  \city{Stanford}
  \country{USA}}
  
  \author{Emily Wen}
\affiliation{%
  \institution{Stanford University}
  \city{Stanford}
  \country{USA}}
  
  \author{Allison Koenecke}
\affiliation{%
  \institution{Microsoft Research and Cornell University}
  \city{Cambridge}
  \country{USA}}
  
  \author{Nikhil Garg}
\affiliation{%
  \institution{Cornell Tech}
  \city{New York City}
  \country{USA}}

  \author{Emma Pierson}
\affiliation{%
  \institution{Cornell Tech}
  \city{New York City}
  \country{USA}}

\renewcommand{\shortauthors}{Zamfirescu-Pereira et al.}

\begin{abstract}
Algorithms provide powerful tools for detecting and dissecting human bias and error. Here, we develop machine learning methods to to analyze how humans err in a particular high-stakes task: image interpretation. We leverage a unique dataset of 16,135,392 human predictions of whether a neighborhood voted for Donald Trump or Joe Biden in the 2020 US election, based on a Google Street View image. We show that by training a machine learning estimator of the Bayes optimal decision for each image, we can provide an actionable decomposition of human error into bias, variance, and noise terms, and further identify specific features (like pickup trucks) which lead humans astray. Our methods can be applied to ensure that human-in-the-loop decision-making is accurate and fair and are also applicable to black-box algorithmic systems. 
\end{abstract}

\maketitle

\section{Introduction}

Recent work in algorithmic fairness has highlighted many of the ways in which algorithms can be biased and error-prone~\citep{mitchell2021algorithmic,chen2020ethical,chouldechova2020snapshot,corbett2018measure}. However, algorithms also provide powerful tools for detecting and dissecting similarly numerous human errors. Understanding patterns of error in human judgment is of interest to a wide range of fields including psychology, computer science, and behavioral economics~\citep{dawes1971,dawes1989, lakkaraju2016confusions,mullainathan2021diagnosing,camerer2019}. Algorithmic and statistical approaches have uncovered systematic human biases---e.g., race or gender biases---in settings including criminal justice~\citep{kleinberg2018human,jung2018omitted,pierson2018fast,pierson2020large,goel2016precinct,lakkaraju2016confusions}, medicine~\citep{pierson2021algorithmic,mullainathan2021diagnosing,lakkaraju2016confusions}, and cultural stereotyping~\cite{garg2018word}. Previous work has also shown the importance of \emph{variance} across decision-makers, in which different decision-makers make inconsistent judgments about similar tasks~\citep{kleinberg2018human,kahneman2021noise}. The use of algorithms to diagnose sources of human error -- whether systematic bias, variance across humans, or unavoidable noise\footnote{Throughout this paper, we use ``bias'' to refer to the case in which humans, on average, misweight features (e.g., race or gender) in making a decision; ``variance'' to refer to inconsistency across human decision-makers when making the same decision; and ``noise'' to refer to irreducible errors made even by Bayes optimal decision-makers. ``Error'' is used as an umbrella term that encompasses all human mistakes, while ``avoidable'' error refers to just bias and variance.} -- has also received increasing attention in the algorithmic fairness community~\citep{abebe2020roles,kleinberg2020algorithms}, in part because humans and algorithms often work together to make decisions,  and so understanding the imperfections of the human in the loop is necessary to achieve overall fairness~\citep{cobbe2021reviewable,valdez2018human,lee2020human,green2019,hilgard2021}.

Here, we develop algorithmic methods to dissect human error in a particular high-stakes task: image interpretation. Humans frequently make important decisions on the basis of images. Clinicians assess x-rays, MRIs, and other image modalities for signs of disease; drivers and pilots respond to fast-changing visual data; online moderators judge whether images are offensive. Understanding patterns of human error in image interpretation has a wide range of applications, including improving training for decision-makers, building algorithmic decision-aids, and deciding when to ask for a second opinion~\citep{bruno2015understanding,raghu2019direct}. However, understanding why and how people err in interpreting images is uniquely challenging. Even defining the salient features in a complex image in an interpretable way~\citep{hollon2020near,wulczyn2021interpretable} is difficult; so is determining how those features influence 1) human decisions and 2) the optimal decisions, and comparing the two in a principled way.  Furthermore, in many datasets the ground truth itself is defined based on human judgments (e.g., in radiology tasks, it is often the consensus opinion of radiologists~\citep{bien2018deep}), and so measuring how humans deviate from ground truth is circular.

We draw on a unique new dataset of human judgments about images to develop a method for understanding human error. In March 2021, \emph{The New York Times} ran a quiz asking respondents to predict whether a Google Street View image came from a neighborhood in which a majority voted for Donald Trump or Joe Biden in the 2020 US election~\citep{trumpbidengeography}. The resulting dataset includes 10,000 neighborhood images, the ground truth for each image (i.e., the true Trump-Biden vote share), and 16,135,392 anonymized individual human predictions of whether a neighborhood voted for Trump or Biden (with more than a thousand human predictions for each image). This dataset is a rich test-bed for methods development because it has 1) an enormous number of human judgments on many unique images; 2) a reasonable prior likelihood that humans perform suboptimally due to stereotypes (e.g., some respondents told journalists that they viewed American flags as predictive of Trump support, but neighborhoods with prominent American flags were actually split evenly between Biden and Trump~\citep{trumpbidenrecap}); and 3) ground truth labels which are derived independently of human judgment (i.e., based on the actual election results), eliminating circularity concerns. While our dataset represents an ideal setting for validating methods, the methods we develop apply to diagnosing human error in image interpretation more generally, as we discuss, as well as to the related task of diagnosing human error in interpreting tabular data and other non-image data modalities.

Using this dataset, we develop a machine learning method to identify when human decision-makers deviate from the Bayes optimal judgment: that is, when they predict ``Trump'' even though the probability that a  neighborhood voted for Biden based on the image is over 50\% (or vice versa). 

This task is not equivalent to identifying \textit{ex post} errors, where human predictions simply disagree with the ground truth. For example, suppose a Street View image happened to capture the only Trump-supporting household (with a prominent Trump flag) in a strongly Biden-leaning neighborhood.  Then, answering ``Trump'' would be Bayes optimal even though it would disagree with the ground truth. This distinction is key in identifying potentially fixable human mistakes and the image features inducing those mistakes, as opposed to cases where the image is uninformative. We make the following contributions:

\begin{itemize} 
\item We propose a method for comparing human decision-making to Bayes optimal decision-making, by first training a machine learning algorithm to estimate the Bayes optimal model and then comparing human decisions to those implied by the estimated Bayes optimal model. We show that even if our estimate of the Bayes optimal model is imperfect, our approach can still provide useful insights into human error as long as our machine learning model adds signal beyond human judgment, a property we verify.
\item We use our method to provide an actionable decomposition of human error into bias, variance, and noise terms by extending a classic decomposition of machine learning model error~\citep{domingos2000unified}. On the Trump-Biden prediction task, we find that noise and variance are larger contributors to human error than is bias.
\item We provide both qualitative and quantitative methods for identifying specific image features which contribute to human error---for example, pickup trucks leading humans to guess ``Trump'' more than is optimal.
\item We analyze, and assess the downsides of, two alternate approaches to diagnosing human error---1) training one model to predict human judgment, training a second model to predict ground truth, and examining deviations between the two models and 2) training a single model to predict the difference between human judgment and ground truth.

\end{itemize} 

While we focus on \emph{human} decision-making in this paper, we note that there is little conceptual difference between assessing, from data, human errors and those of black-box technical systems where only the system's input and output is known. Our method only makes use of the input image, humans' binary judgements, and ground truth---and so could be used, for example, to audit black-box third-party vision APIs which only output binary decisions. As such, our work contributes to a long line of work using algorithmic approaches to audit other computational systems~\citep{lurie2019opening,chen2018investigating,ali2019discrimination,datta2014automated,sweeney2013discrimination,bolukbasi2016man,caliskan2017semantics,buolamwini2018gender,koenecke2020,sandvig2014auditing,abebe2019using,abebe2020roles,metaxa2021auditing}.

\section{Related work} \label{sec:related_work}

Measuring and describing error---and in particular, bias and variance---in human decision-making has been a problem of interest to social scientists for decades, ranging from theoretical models of racial, gender, and other types of discrimination~\citep{becker2010economics} to cognitive heuristics which are employed under uncertainty~\citep{tversky1974judgment}. ~\citet{kahneman2011thinking} provides a recent review of common biases in human decision-making, and ~\citet{kahneman2021noise} addresses the importance of variance, in which humans make inconsistent judgments about similar problems. Our work is motivated by, and builds on, this literature by using modern machine learning approaches to decompose human error into bias, variance, and noise terms, and then to explain image-specific causes. Below, we summarize the algorithmic communities closest to our work.

\paragraph*{Algorithmic approaches to measuring error in human decision-making.}
In recent years, there has been increasing interest in the algorithmic fairness community in using computational and algorithmic approaches to measure error in \emph{human} decision-making, as part of a broader recognition that such approaches can serve a useful \emph{diagnostic} function in precisely understanding and measuring social problems~\citep{abebe2020roles,garg2018word}. Several observations motivate this interest. First, much prior work has argued that algorithmic decision-making is (theoretically if often not practically) more transparent than human decision-making, allowing algorithms to serve as ``discrimination detectors''~\citep{kleinberg2020algorithms,kleinberg2018discrimination,mullainathan2019biased}. Second, algorithmic tools are often designed to be used by humans, so understanding the imperfections of the human in the loop is necessary to ensure the system as a whole is fair~\citep{cobbe2021reviewable,valdez2018human,lee2020human}. Failing to account for human biases can produce algorithms which perform well on retrospective data but yield unexpected or pernicious effects when they are actually used by human decision-makers~\citep{stevenson2021algorithmic}. 

Motivated by these observations, there have been numerous examples of using algorithmic approaches to uncover systematic human biases. Algorithms have been used to diagnose broader cultural biases and stereotypes, often through use of word embeddings~\citep{garg2018word,caliskan2017semantics,bolukbasi2016man,demszky-etal-2019-analyzing}; this work differs from ours because it studies broader cultural trends but not specific human errors in decision-making. Closer to our own work is the use of algorithms to study human errors in decision-making settings. For example, in criminal justice~\citep{kleinberg2018human,jung2018omitted,pierson2018fast,pierson2020large,goel2016precinct,lakkaraju2016confusions}, algorithms have been used to diagnose human error in bail decisions~\citep{kleinberg2018human,lakkaraju2016confusions} and stop-and-frisk~\citep{goel2016precinct,jung2018omitted}, among other settings. In medicine, algorithms have been used to identify human errors in diagnosing pain~\citep{pierson2021algorithmic}, diagnosing heart attacks~\citep{mullainathan2021diagnosing}, and prescribing asthma treatments~\citep{lakkaraju2016confusions}. Algorithms have also been used to diagnose and dissect human error in games like chess~\citep{anderson2017assessing,mcilroy2020learning,mcilroy2020aligning}.

Our work builds on this literature by developing and evaluating machine learning methods for diagnosing how humans err in analyzing \emph{image data specifically}; in contrast, the work above focuses on tabular data. As discussed above, analyzing human decisions made from image data poses unique challenges. As a representative illustration of how methods for tabular data do not easily transfer to image data, consider the work of~\citet{jung2018omitted}, who convincingly demonstrate bias in police search decisions in New York City using tabular data. Based on historical prior knowledge, they specifically assess racial bias, which they quantify by measuring the racial disparities in searches which remain when controlling for (a model's estimate of) a pedestrian's objective risk of carrying contraband. Their approach of assessing human decisions relative to objective risk is conceptually similar to ours (see ~\citet{mullainathan2021diagnosing} for another example of this approach). However, their method cannot be directly applied to our setting because it relies on tabular data with 1) clearly interpretable features and 2) an \emph{a priori} understanding of which features are likely to contribute to human biases (in their setting, race); in contrast, with image data, we have neither of these things. While we confront the unique challenge of image data, many of our methods and observations also apply to tabular data because it is an easier task.

\paragraph*{Algorithmic descriptions of human decision-making.} Beyond specifically diagnosing human error, algorithmic approaches have also been used to describe human decision-making more broadly~\citep{Agrawal8825,hilgard2021}. Our approach of analyzing human behavior via a residual with a machine learning model is somewhat similar to that of \citet{Agrawal8825}; however, while they use a neural network to smooth  high-dimensional noise in empirical human decisions when \textit{predicting human decisions}, we use one as a proxy for the Bayesian optimal decision when \textit{analyzing human error}. 

\paragraph*{Human-algorithmic collaborations.} There is substantial work on creating algorithms which can learn to \emph{complement} humans~\citep{tan2018,green2019,dearteaga2020,hilgard2021}, for example, by learning to defer to a human expert when the human will achieve better performance~\citep{wilder2020learning,madras2017predict,mozannar2020consistent} or by providing automated assessments for human experts to consider in making decisions~\citep{groh2022deepfake}. Such approaches often implicitly model human error. Our work differs in that it focuses on explicitly describing human error, not in creating algorithms which implicitly model human error while learning to complement humans.

\paragraph*{Computer vision interpretability.} We describe human and optimal decision-making from images using convolutional image models, and consequently rely on methods for interpreting these models. There is a wide literature on such methods: see~\citet{zhang2018visual} for a review. In our primary results, we make use of occlusion mapping~\citep{zeiler2014visualizing}, a commonly used technique which identifies regions of the image which influence a model's prediction by determining how much the prediction changes in response to masking out regions of the image.

\section{Problem setup} \label{sec:setup}

\subsection{Data}
\label{sec:dataset}

\paragraph{The New York Times quiz dataset.} Our main dataset consists of the 10,000 neighborhood images available in the \emph{New York Times} quiz, which we partition into 5,000 training images; 1,500 validation images; 1,500 images which we use as a preliminary test set while conducting experiments for this paper; and a holdout test set of 2,000 images which we use only to generate the final results for this paper to minimize overfitting~\cite{kleinberg2018human,pierson2021algorithmic}. All our main results are reported on this holdout test set. Each image is retrieved from Google Street View as a composite of four individual views for the same location at viewing angles of $0^\circ$, $90^\circ$, $180^\circ$, and $270^\circ$, such that the stitched composite image approximates what human respondents could see in the \emph{New York Times} quiz.  Each image is linked to the ground truth Trump-Biden vote share in that neighborhood.\footnote{We use ``neighborhood'' throughout to refer to \emph{electoral precinct}, the most granular area in the United States for which election results are publicly available. When we say that a neighborhood voted for a candidate, we mean that a majority of voters in that neighborhood -- who voted for either Trump or Biden -- voted for that candidate.} Additionally, each neighborhood image has an average of 1,614 corresponding human predictions of whether a neighborhood voted for Trump or Biden.\footnote{While we are able to identify which predictions came from the same \emph{New York Times}-identified human---who each on average made predictions on 10.33 images in our training data---we do not know the order in which the respondent made their predictions on different images. As such, we cannot assess the effect of feedback received by each respondent over their sequence of predictions. Furthermore, \emph{The New York Times} did not collect any additional information on respondents, such as demographic information, location, or IP address. Thus, we cannot assess how performance or human error varies by such covariates.}

\emph{Augmenting the quiz dataset.} To increase the amount of data we have to train our models, we collect an additional external dataset of 52,025 Google Street View images linked to election results (but no human judgments), using a sampling method similar to that of \emph{The New York Times}. See Appendix \ref{sec:data_augmentation} for details.

\begin{figure}[tb!]
  \centering
  \includegraphics[width=3in]{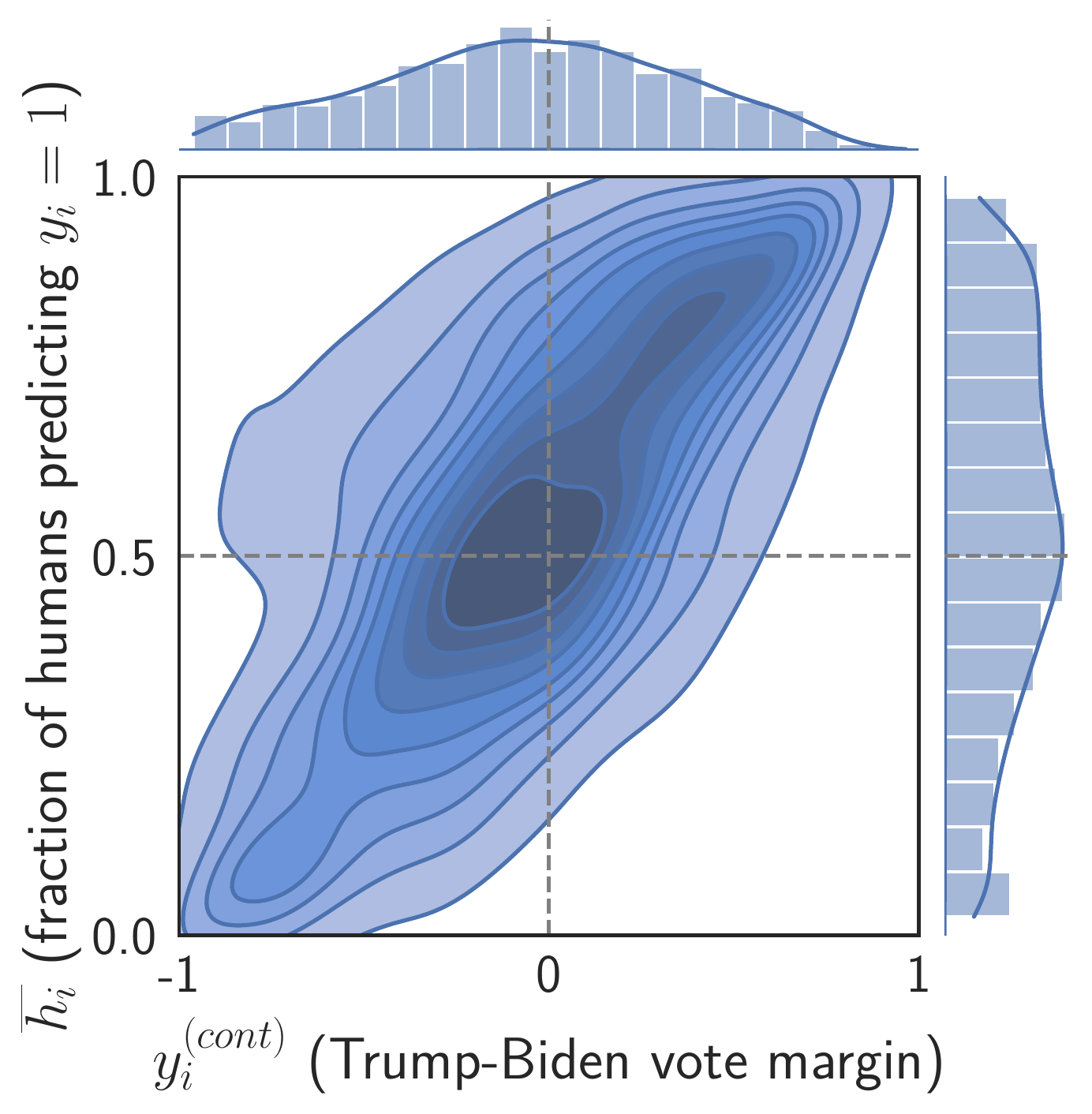}
  \caption{Joint histogram of $\gtcont$, the true Trump-Biden vote margin (x-axis), against $\hagg$, the fraction of humans who guessed ``Trump'' (y-axis). Humans deviate substantially from both omniscience (which would imply a threshold function at $\gtcont = 0$) and perfect agreement with each other (which would imply $\hagg \in \{0, 1\}$). However, note that this plot is not enough to conclude that humans in aggregate are making avoidable mistakes -- it could be that the images are uninformative, and so the errors are due to noise that even a Bayes optimal decision-maker would make. Our methods are designed to separate such noise from avoidable human errors.}
  \label{fig:scatterplot-hagg-vs-gt}
\end{figure}

\subsection{Notation} Throughout, $i$ indexes neighborhood images and $j$ indexes human judgments about each image. For each image $\image$, we have human judgments of whether the majority vote in the associated neighborhood  was for Trump or Biden, where $\hind \in \{0, 1\}$ denotes individual human judgments; a 0 indicates Biden, and 1 indicates Trump. We use $\hagg \in [0, 1]$ to denote the mean human judgment for each image (i.e., the fraction of people who indicated Trump for that image). We also have binary ground truth, $\gt \in \{0, 1\}$, indicating the true majority vote for the corresponding neighborhood; it is also useful to refer to the continuous vote share difference (the fractional Trump vote share minus the fractional Biden share), denoted $\gtcont \in [-1, 1]$, with $\gt = \mathbbm{1}[\gtcont > 0]$. We report additional statistics for ground truth and human judgment in Table \ref{tab:quiz_stats}.

Figure \ref{fig:scatterplot-hagg-vs-gt} plots $\gtcont$ against $\hagg$, showing that, while human judgment is correlated with true election outcomes, humans are far from omniscient and frequently disagree on each image. However, deviations from $\gtcont$ are not enough to conclude that humans are making \emph{avoidable} mistakes -- it could be that the images are uninformative about election outcomes, and so the errors are unavoidable ones that even a Bayes optimal decision-maker would make.

\begin{table*}[h]
    \centering
    \begin{tabular}{l|c|c|c|c}
        & Train & Validation & Test \\
        \hline
        \# of images & 5,000 & 1,500 & 2,000 \\ 
        \# of human responses & 8,064,385 & 2,420,386 & 3,229,708 \\
        \hline
        Accuracy of individual human & 0.629 & 0.635 & 0.627 \\
        Accuracy of aggregate human & 0.724 & 0.727 & 0.707\\
        \hline
        Responses per image (mean) & 1,613 & 1,614 & 1,615 \\
        Responses per image (median) & 1,611 & 1,613 & 1,614 \\
        Responses per image (std) & 68.04 & 72.80 & 62.05 \\
        \hline
        Responses per user ID (mean) & 10.33 & 3.60 & 4.51 \\
        Responses per user ID (median) & 8 & 3 & 3 \\
        Responses per user ID (std) & 23.53 & 7.65 & 9.88 \\
        \hline
        Fraction of images where $\gt = 1$ & 0.458 & 0.444 & 0.449 \\
        Fraction of images where $\hagg > 0.5$  & 0.563 & 0.557 & 0.567 \\
        Fraction of responses where $\hind = 1$ & 0.537 & 0.535 & 0.538
        
    \end{tabular}
    \caption{Descriptive statistics for the \emph{New York Times} train, validation, and held-out test datasets. The differences across  datasets for responses per user ID are expected, as the training set has approximately triple the number of images as do the other sets. All our main results are reported on the holdout test set (final column).}
    \label{tab:quiz_stats}
\end{table*}

To model such a Bayes optimal decision-maker, we let $p(\gt=1|\image)$ denote the probability that the neighborhood corresponding to a certain image voted for Trump; i.e., how often images that look like $\image$ correspond to a neighborhood where the majority voted for Trump. For example, suppose 80\% of neighborhoods with pickup trucks voted for Trump; then, $p(\gt=1|\image = \text{\textit{Neighborhood with pickup truck}})$ = $0.8$.   

\subsection{Task} A Bayes optimal decision-maker constrained to make binary judgements about each image should predict $\gt=1$ if and only if $p(\gt=1|\image) > 0.5$. Here, we aim to quantify to what extent human decision-makers deviate from Bayes optimality and identify what features of the images lead to these errors. This is a task of primary interest because it quantifies to what extent, and why, humans are making \emph{avoidable} (and potentially fixable) errors in interpreting images. However, this task is challenging for several reasons: 1) the Bayes optimal decision, and in particular the probability $p(\gt=1 | \image)$, is not directly observable in the data; 2) we also do not observe individual human estimates of $p(\gt=1 | \image)$, just their binary decisions; and 3) defining salient features in the images is difficult. 

Our methods, described next, are designed around these challenges. We note that while our notation and descriptions are particular to our dataset, these characteristics are common for many settings in which humans make decisions using images.

\section{Method}  \label{sec:method}

Our main approach to characterizing human error examines how humans deviate from the Bayes optimal decisions implied by $p(\gt=1|\image)$. In \Cref{sec:estimating_bayes_optimal_classifier} we describe how we fit a model to estimate $p(\gt=1|\image)$ and provide evidence of the quality of the estimate. In \Cref{sec:decomp}, we show how our estimate of $p(\gt=1|\image)$ can be used to decompose human error into bias, variance, and noise terms. Finally, in \Cref{sec:whybias}, we show how our estimate of $p(\gt=1|\image)$ can be used to identify specific image features which lead humans astray.

\subsection{Estimating and plotting $p(\gt=1|\image)$} 
\label{sec:estimating_bayes_optimal_classifier}

\paragraph{Training a model to estimate $p(\gt=1|\image)$} We estimate this model in two steps. First, we train a deep learning model to estimate $p(\gt=1|\image)$ on the large external dataset collected as described in Section \ref{sec:dataset} (see \Cref{sec:training_image_models} for model training details; we verify that training a model on the large external dataset yields slightly superior performance to training only on the smaller \emph{New York Times} dataset). We use $\fhatext$ to denote our deep learning model's estimate of $p(\gt=1|\image)$. Second, using the \emph{New York Times} training and validation datasets, we fit a simple logistic regression to estimate $ p(\gt=1|\image)$ using both the deep learning model's prediction $\fhatext$ and the aggregate human judgment $\hagg$ as features. In other words, we estimate $p(\gt=1|\image) = \text{sigmoid}(\alpha + \beta_1 \fhatext + \beta_2 \hagg)$.\footnote{We use a model with this simple parametric form because we find no evidence that more complex models---e.g., with interaction terms---improve our predictions.} The predicted probabilities from this logistic regression model constitute our final estimate of $p(\gt=1|\image)$; we use $\fhat$ to refer to this estimate.

This two-stage procedure---first fitting a deep learning model to estimate the probability an image voted for Trump, and then estimating a logistic regression which combines the model output with human judgment---has three benefits:

\begin{enumerate} 
\item Inspecting the logistic regression coefficients allows us to \emph{verify}, rather than \emph{assuming}, that our model $\fhatext$ truly identifies ground-truth relevant signal that humans miss. If $\fhatext$ provided no additional signal beyond human judgment $\hagg$, the coefficient on $\fhatext$ in the fitted logistic regression model would be zero. Instead, the coefficient on $\fhatext$ is 22.2 (95\% confidence interval, 20.5---24.0). The large and significant coefficient on $\fhatext$ indicates that the machine learning model is indeed detecting ground-truth relevant signal that humans miss and is thus useful for diagnosing human error.
\item Our goal is to estimate the Bayes optimal model as accurately as possible, so we should use all features derivable from the image, including human judgments based on the image. There is no guarantee that the deep learning model will capture all the signal on the image, given that it is trained on an external and finite dataset---although we do provide evidence below that our model is trained on a sufficiently large dataset for performance to level off, suggesting it approaches optimality. Substantiating this reasoning, the coefficient on $\hagg$ in our logistic regression is 2.0 (95\% confidence interval, 1.5---2.4). The statistically significant coefficient on $\hagg$, though much smaller than that on $\fhatext$, indicates that humans in aggregate also pick up at least some signal that $\fhatext$ misses---likely because $\fhatext$ is trained on an finite external dataset which may not totally match the \emph{New York Times} distribution. Combining both human judgment and the algorithmic prediction thus yields the best approximation of the Bayes optimal decision, which is our goal. 
\item Finally, as we discuss below, this two-stage estimation procedure will yield a useful lower bound on the magnitude of human bias even if the deep learning model we fit does not capture all the signal in the image.
\end{enumerate}

Overall, a key strength of our two-stage approach is that it does not rely on being able to learn a machine learning model which perfectly estimates $p(\gt=1|\image)$ in order to provide useful insight into human error --- often an impossible desideratum in small-data regimes. Rather, it merely requires that the machine learning model adds \emph{some signal} beyond that captured in human judgment --- something we can directly verify through logistic regression. Consistent with this, the accuracy of our final model $\fhat$ in predicting ground truth $\gt$ (75.1\%) exceeds the accuracy of aggregate human judgment $\hagg$ (70.7\%) or individual human judgment $\hind$ (62.7\%).

\paragraph{Assessing the quality of the estimate of $p(\gt=1|\image)$}

We verify two properties of our estimate of $p(\gt=1|\image)$. First, we show that it is calibrated (\Cref{fig:calibration-plot}) by comparing the model's predicted probabilities to the true fraction of positive examples for groups of observations binned by predicted probability, a standard check. Second, we verify that the machine learning model $\fhatext$ which forms a component of our final estimate of $p(\gt=1|\image)$ is trained on a sufficiently large dataset for model performance to level off, suggesting that we have enough training data that model performance is reasonably close to optimal: in Figure \ref{fig:downsample_plot}, we plot model performance with training sets of various sizes, showing that model performance levels off prior to our training set size. As we discuss below, one advantage of our two-stage approach is that it can still yield useful insights into human error even if $\fhatext$ is not completely optimal, as long as it adds signal beyond human judgment. Still, our approach will have more power to detect human error if the machine learning model it relies on is reasonably close to optimal performance, which is why we perform this check.

\begin{figure*}[tb]
  \centering
    \centering
  \subcaptionbox{ \label{fig:sfig1}}{%
  \includegraphics[width=0.5\linewidth]{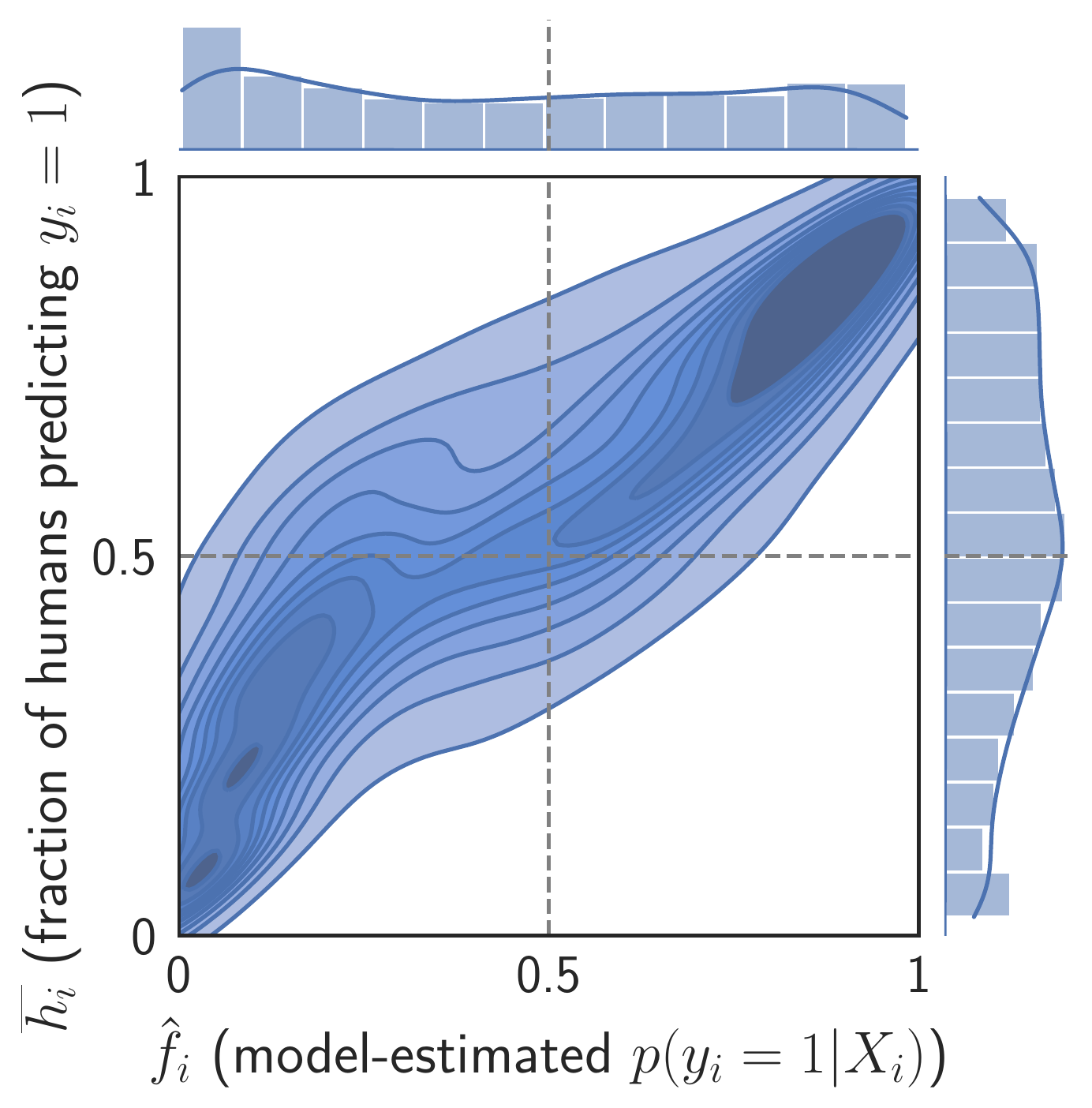}%
}\hfil
\hfil
\subcaptionbox{\label{fig:sfig3}}{%
  \includegraphics[width=0.45\linewidth]{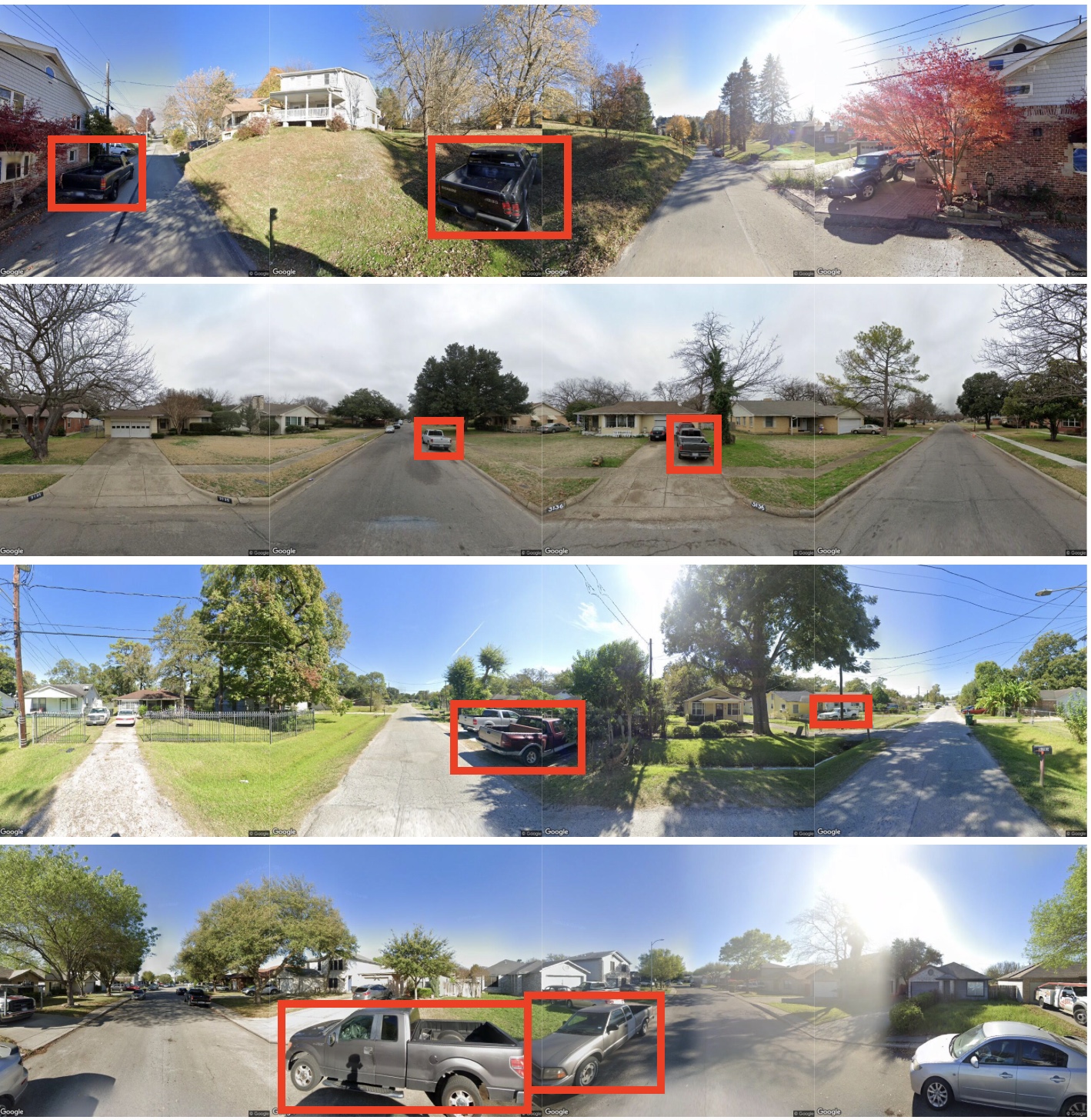}}%

  \caption{\textbf{(a)} Plotting $\fhat$, the model-estimated $p(\gt=1|\image)$, (x-axis) against aggregated human judgment $\hagg$ (y-axis) reveals that human judgment deviates substantially from Bayes optimality, which would produce a threshold function at $\fhat = 0.5$. Instead, even for images where $fhat$ is only 25\%, far more than 50\% of respondents predicted Trump in some cases. Unlike~\Cref{fig:scatterplot-hagg-vs-gt}, this figure establishes that humans are making \emph{avoidable} errors -- ones not made by an approximately Bayes optimal decision-maker shown the same images the humans as human respondents. \textbf{(b)} 15 of the 21 images where humans most incorrectly skew towards Trump have pickup trucks (red bounding boxes), illuminating a source of human bias; we show 4 examples here; all 21 images also feature wide regions of open sky unobstructed by buildings. These images are identified by filtering for images with estimated $p(\gt=1|\image) < 0.2$ (Biden-leaning) and $\hagg > 0.6$ (Trump-leaning)---i.e., images in the top left region of (a). Underlying street view images \copyright Google.}
\end{figure*}

\paragraph{Assessing how $\hagg$ deviates from decisions implied by $\fhat$} Next, we conduct a preliminary assessment of how human decision-making compares to what we would expect if humans were Bayes optimal. \Cref{fig:sfig1} plots $\fhat$, the model-estimated $p(\gt=1|\image)$, against $\hagg$. The relationship differs considerably from what we would expect if humans were Bayes optimal, in which case we would see a threshold function $\hagg = \mathbbm{1}[\fhat > 0.5]$. This implies that human decision-making is imperfect. Note that, unlike \Cref{fig:scatterplot-hagg-vs-gt}, this figure suggests that humans are making \emph{avoidable} errors---i.e., decisions which deviate from those of an estimated Bayes optimal decision-maker which has access to just the same images the humans do. 

As a preliminary analysis of human error, we manually inspect individual images where $\hagg$ deviates particularly dramatically from decisions implied by $\fhat$. Note that because $\fhat$ is learned from both $\hagg$ and $\fhatext$, if $\fhatext$ added no additional signal above $\hagg$ for predicting ground truth, $\fhat$ and $\hagg$ would be perfectly correlated, and there would be no dramatic deviations to examine at all; the presence of such deviations only occurs because $\fhatext$ does indeed add additional signal. We examine images where $\fhat$ has high confidence that the neighborhood voted for Biden, but humans disagree---i.e., images in the top left corner of \Cref{fig:sfig1}. (We set cutoffs at $\fhat<0.2$ and $\hagg > 0.6$, but our results are not sensitive to these thresholds.) 21 of the holdout test set images meet these criteria. Importantly, we find that of these images, only 19\% in fact voted for Trump, indicating that $\fhat$ is correct that these images are likely Biden neighborhoods and humans are in fact making avoidable errors. These images disproportionately have pickup trucks (trucks appear in 71\% of these images as opposed to in 41\% of the holdout test set as a whole\footnote{In a random sample of 100 images from our holdout test set, manually inspected, we counted 41 images containing pickup trucks.}), indicating that humans believe trucks predict Trump more often than they really do (Figure~\ref{fig:sfig3}). This observation is indeed consistent with what \emph{The New York Times} heard from some survey respondents, who in interviews said they believed that ``pickup trucks were clear indications of a community’s more conservative politics.''
All 21 images in this set also have wide regions of open sky unobstructed by buildings, another source of bias we discuss further in \Cref{sec:residual-interpretation}. 

We identify no images where humans display the opposite bias---where the model is very confident that a neighborhood voted for Trump, $\fhat>0.8$, but humans disagree, $\hagg<0.4$---indicating an interesting asymmetry in human judgment. Consistent with this, \Cref{tab:quiz_stats} shows that humans are slightly miscalibrated: they think neighborhoods vote for Trump more often than they really do\footnote{Because we are interested in assessing all human biases, we do not calibrate human decisions prior to assessing them. However, calibrating $\hagg$ by choosing a threshold such that $\hagg$ classifies the correct number of images as Trump only increases its accuracy slightly (from 70.7\% to 72.1\%) likely because human decisions near the boundary are very noisy.}. Having established that simply examining the plot of how $\hagg$ deviates from decisions implied by $\fhat$ can provide interesting insights into human error, we explore methods for more systematically decomposing this error below.

\subsection{Bias-variance-noise decomposition of human error} 
\label{sec:decomp}

\begin{figure}[tb!]
  \centering
  \includegraphics[width=3in]{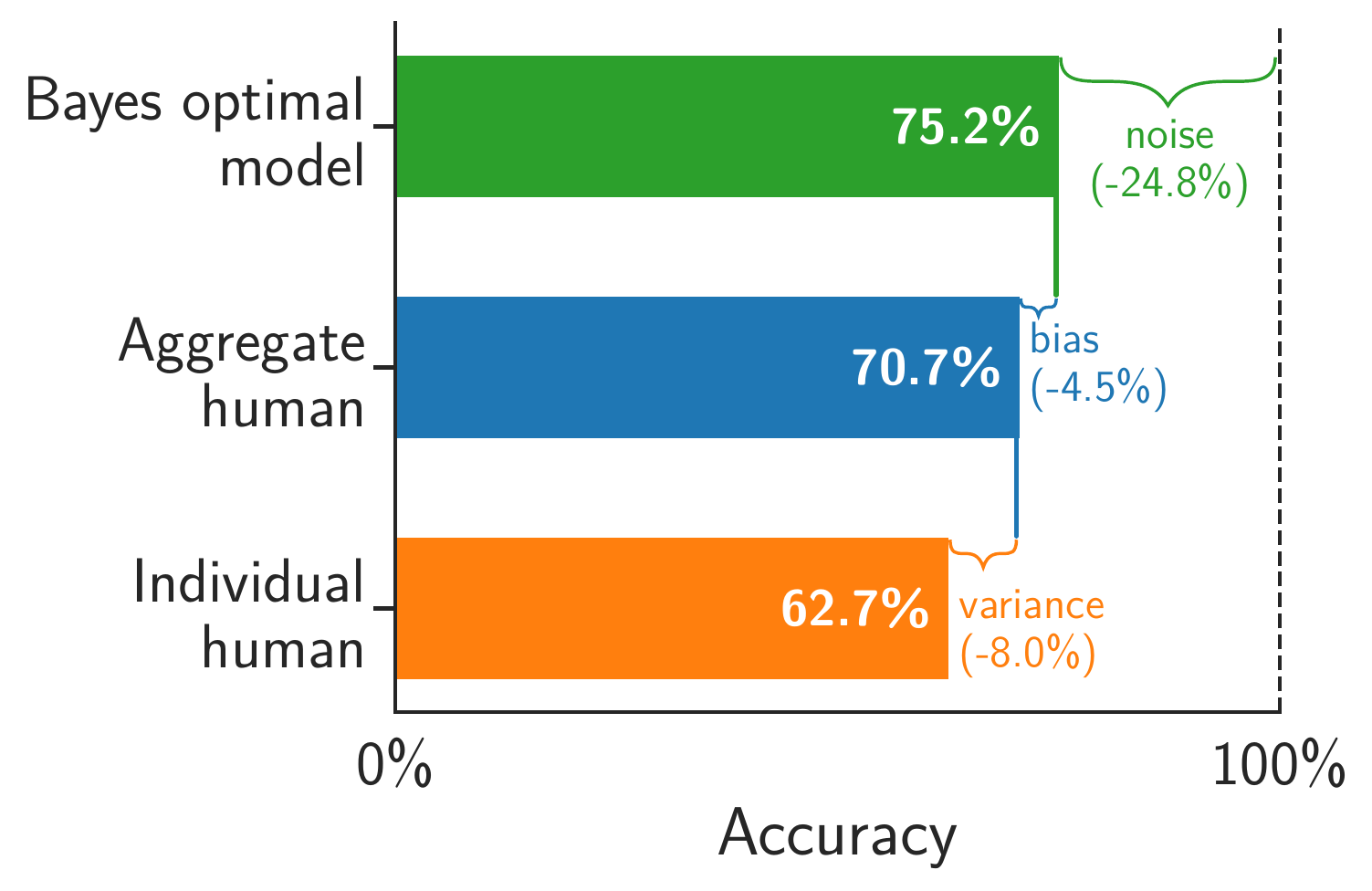}
  \caption{Sources of human error are decomposed into noise, bias, and variance terms by plotting the accuracy of our estimated Bayes optimal model, of aggregated human judgments, and of individual humans. (Pairwise differences between bars denote percentage point differences in accuracy.) As discussed in \Cref{sec:decomp}, \textit{noise} is irreducible error that even the estimated Bayes optimal model cannot avoid; \textit{bias} is additional error made by the aggregate human judgment for each image; and \textit{variance} is the additional error due to disagreements between humans judging the same image. }
  \label{fig:sfig2}
\end{figure}

Estimating $p(\gt=1|\image)$ also allows us to decompose human error into three sources---bias, variance, and noise---inspired by a classic decomposition for machine learning classifiers~\citep{domingos2000unified}, and related to past work which seeks to assess both bias and variance in human decisions~\citep{kleinberg2018human}. Prior to introducing our decomposition in the human decision-making setting, we briefly review the original decomposition in the machine learning setting~\citep{domingos2000unified}. Given a machine learning algorithm for learning a classifier, a training set of a fixed size, and a set of covariates, the goal of the decomposition is to assess why the machine learning algorithm performs imperfectly. ~\citep{domingos2000unified} defines the \emph{main prediction} as the aggregated prediction (e.g., majority vote in the case of zero-one loss) of classifiers fitted on different draws of the training set. Given this, the \emph{bias} of the algorithm is the loss of the main prediction relative to the Bayes optimal prediction; the \emph{variance} is the average loss of classifiers learned from individual training sets relative to main prediction; and the \emph{noise} is the loss of the Bayes optimal classifier. In other words, the bias captures accuracy loss due to the inability of the model family to capture the true Bayes optimal model; the variance captures accuracy loss due to random variation across classifiers fitted on a finite train set; and the noise captures accuracy loss due to intrinsic unpredictability of the outcome from the features.

Our extension of this formalism to human decision-making is intuitive. We define the \emph{main prediction} for humans as the binarized majority vote for each image, $\mathbbm{1}[\hagg > 0.5]$, and conceptualize each human decision-maker as single fitted classifier from the ``human model class''. In Figure \ref{fig:sfig2} we plot (1) the accuracy of the estimated Bayes optimal model $\fhat$; (2) the human accuracy if every human agreed with the main prediction on each image; and (3) the accuracy of individual human predictions. The difference between (1) and perfect performance is the accuracy loss due to noise; the difference between (1) and (2) is the accuracy loss due to human bias; and the difference between (2) and (3) is the accuracy loss due to human variance. The accuracy losses due to variance (8.0 percentage points) and noise (24.8 percentage points) exceed those due to bias (4.5 percentage points). Thus, in this setting, our results establish that much of the error on the binary prediction task is unavoidable, and made even by our estimated Bayes optimal model: it is simply difficult to predict a neighborhood's election outcomes from a single Street View image.
The fact that the accuracy of the aggregated human judgment $\hagg$ exceeds that of individual human judgments is consistent with previous work demonstrating wisdom of crowds~\cite{galton1907vox}. If this were a real-world decision-making task, our results imply that having several humans judge each image would yield a considerable improvement over individual judgments, approaching the performance of a machine learning model. We note that if our model $\fhat$ fails to capture the Bayes optimal model, we will overestimate the accuracy loss due to noise, and underestimate the loss due to human bias. Thus, our approach provides a useful \emph{lower bound} on the magnitude of human bias even if our estimated model is not optimal.

More broadly, we believe that our decomposition provides an actionable heuristic for assessing and improving decision-making processes in a wide variety of settings. For example, if doctors on a medical image classification task mainly lose accuracy due to bias, we may wish to consider retraining them or replacing them with an automated system; if they are accurate in aggregate but individually high-variance, we may need to solicit second opinions; and if the images themselves are noisy, we may need an alternate diagnostic modality.

\subsection{What image features influence human judgment beyond the objective probability $p(\gt=1|\image)$?} 
\label{sec:whybias}

Our manual inspection of images (\Cref{sec:estimating_bayes_optimal_classifier}) shows that deviations between $\hagg$ and $\fhat$, our estimate of $p(\gt=1|\image)$, can provide insights about image features which lead humans astray, like pickup trucks (Figure \ref{fig:sfig3}).

We now assess whether we can \emph{systematically predict} from the image when $\hagg$ will deviate from what we would expect given our estimate of $p(\gt=1|\image)$: if there are image features that produce such systematic deviations, it suggests that humans are influenced by these features beyond what $p(\gt=1|\image)$ would justify. For example, consider our running example of pickup trucks, and suppose there is a pair of images with the same $p(\gt=1|\image)$---one with a pickup truck, and one without. If $\hagg$ for the image with the truck is greater than $\hagg$ for the image without the truck, this disparity is not justified by the objective probability of the image $p(\gt=1|\image)$: perhaps $\hagg$ for the truck image is too high, or $\hagg$ for the non-truck image is too low, but we can be sure that humans have a Trump-truck association beyond that justified by $p(\gt=1|\image)$. (Note that there may be some justified association between pickup trucks and probability that the neighborhood voted for Trump; however, this justified association would be captured in $p(\gt=1|\image)$).

\begin{figure*}
  \centering
  \subcaptionbox{ \label{fig:polyline-residual-performance}}{%
  \includegraphics[width=0.65\linewidth]{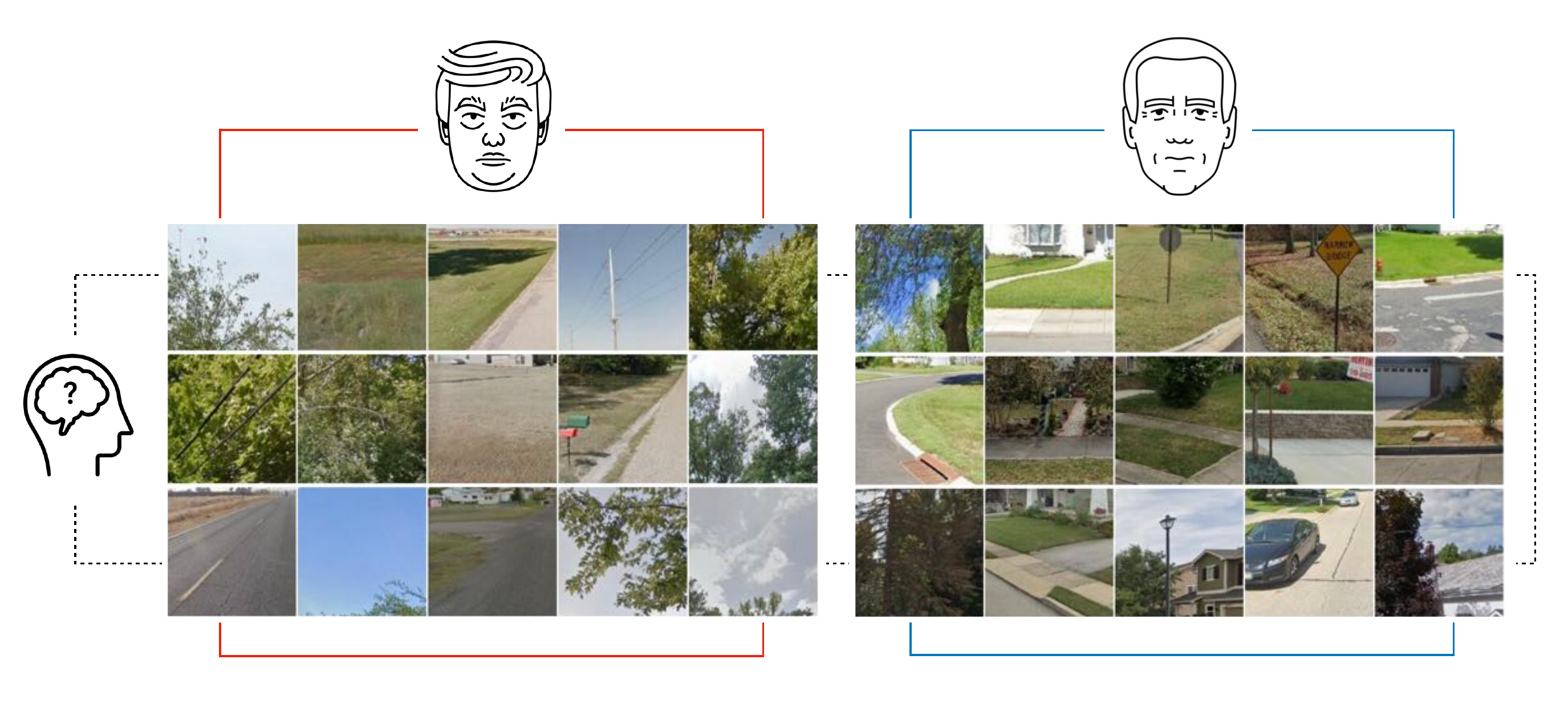}%
}\hfil
\hfil
\subcaptionbox{\label{fig:density_error_plot}}{%
  \includegraphics[width=0.3\linewidth]{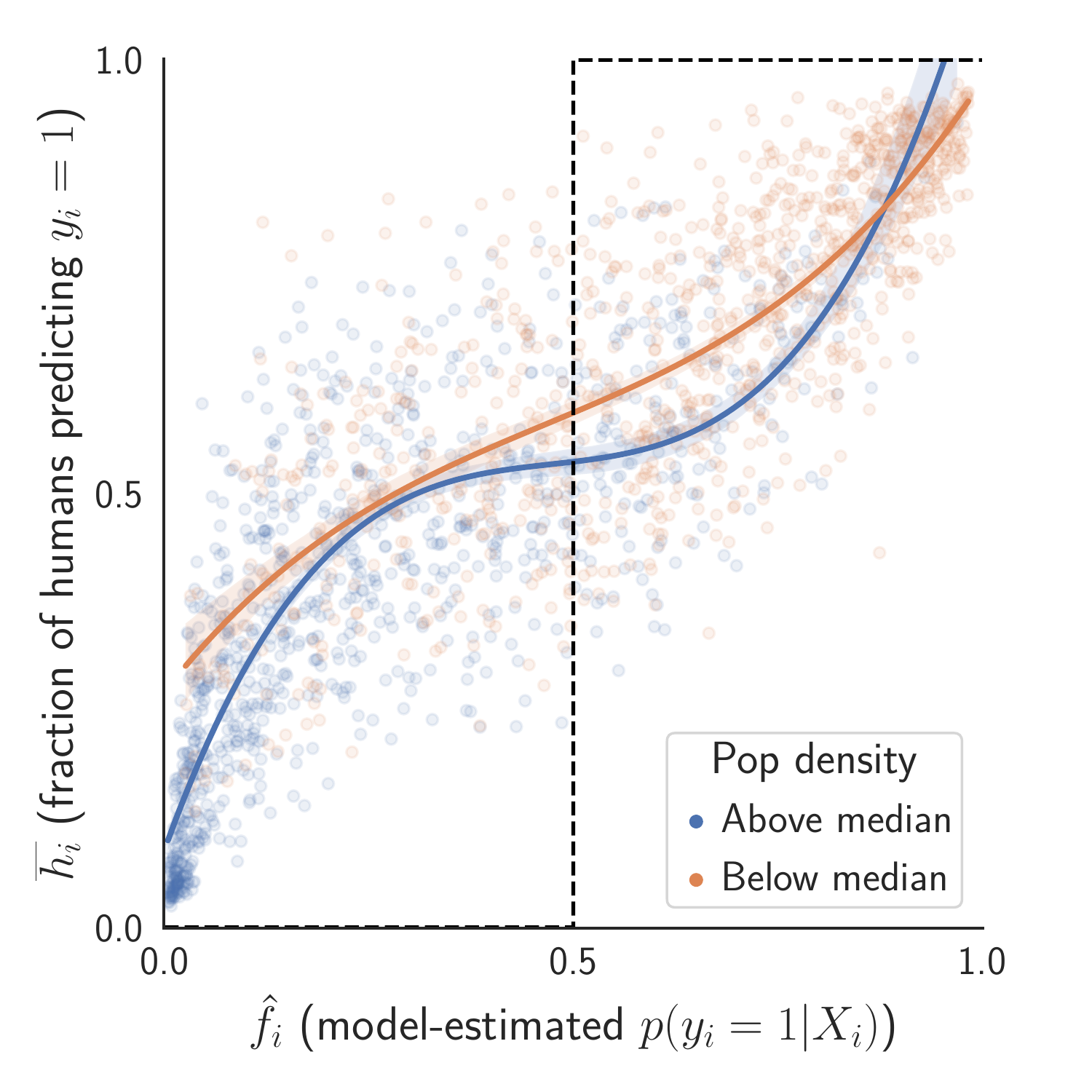}%
}

  \caption{Density influences human decision-making beyond what is justified by $\fhat$. \textbf{(a)} Image patches that caused the most shift in the residual model error towards Trump (left patches) or Biden (right patches). Patches of sky and other markers of low density, like road edges without curbs, on the left suggest that human respondents are more influenced towards Trump by features suggesting low population density than $\fhat$ can explain. To further substantiate this observation, \textbf{(b)} shows how $\hagg$ varies as a function of $\fhat$ for neighborhoods with higher than median population density (blue line) and lower than median population density (orange line). The orange line is higher than the blue line, indicating that, controlling for $\fhat$, humans are more likely to think that less dense neighborhoods are Trump neighborhoods. The dotted line shows the Bayes optimal decision boundary. Underlying street view images \copyright Google.} 
 
\end{figure*}

To formalize this intuition, we train a second model to predict, from the image, how much $\hagg$ deviates from what we would expect given $p(\gt=1|\image)$. As before, we approximate $p(\gt=1|\image)$ with $\fhat$. We do this in three steps: 

\begin{enumerate} 
\item We first flexibly capture how $\hagg$ varies as a function of $\fhat$ by fitting a cubic polynomial $\hagg = m(\fhat)$. $m(\fhat)$ corresponds to what the human average $\hagg$ tends to be for an image with estimated probability $\fhat$.
\item We then define the \emph{residual}  $\residual = \hagg - m(\fhat)$: that is, the portion of $\hagg$ that differs from how the aggregate human judgment tends to behave for images with the corresponding estimated probability $\fhat$. If there are image features that predict this residual, this suggests that those image features lead to inconsistent human judgments about images with the same estimated probability $\fhat$. We note that because we estimate $\fhat$ using both $\hagg$ and $\fhatext$, if $\fhatext$ added no additional signal above $\hagg$ for predicting ground truth, $\fhat$ and $\hagg$ would be perfectly correlated and the residual would be uniformly zero. Thus, in trying to predict a non-zero residual, we are attempting to predict signal which truly arises from the fact that the machine learning model $\fhatext$ identifies ground-truth relevant signal which humans miss. 
\item To search for the image features which predict the residual, we train a neural network to predict the residual from the image: $\hat \residual = g(\image)$ (see Appendix \ref{sec:training_image_models} for model training details). 
\end{enumerate} 
We find that the neural network is able to achieve statistically significant signal for predicting the residual from the image (Spearman correlation between true and predicted residual, $0.534$; $p<0.001$).
This statistically significant correlation shows that $\residual$ is systematically predictable from the image, indicating that there are image features which cause humans to deviate systematically from consistent responses to the estimated objective probability $\fhat$.

\paragraph{Interpreting the residual model.} \label{sec:residual-interpretation} To identify the specific image features which cause this systematic deviation in the residual, we use occlusion mapping~\cite{zeiler2014visualizing} to interpret the fitted residual model. Specifically, we identify the image patches which most change the residual model's predictions when they are masked out (Appendix~\ref{sec:image_patches}). The results of this analysis are shown in Figure~\ref{fig:polyline-residual-performance}, illustrating that patches of sky and other features indicating low population density push the residual model in the Trump direction: in other words, human respondents are more swayed towards Trump by visual indicators of low population density than the estimated objective probability can explain.\footnote{The open sky patches in \Cref{fig:polyline-residual-performance} are consistent with the open skies seen in individual images in \Cref{fig:sfig3}, indicating that the manual inspection of individual images is yielding conclusions consistent with the quantitative residual analysis.} As further evidence, the correlation between $\hat \residual$ and log population density is also negative (Spearman $r$ of $-0.234$, $p<0.001$): \Cref{fig:density_error_plot} illustrates that, controlling for the estimated objective probability, humans think that denser neighborhoods are more likely to be Biden neighborhoods.
Overall, this analysis shows that humans are swayed by population density beyond what our estimate of $p(\gt=1|\image)$ can explain: given two images with identical estimated $p(\gt=1|\image)$, humans will be more likely to think the denser neighborhood voted for Biden. Interviews conducted by \emph{The New York Times} confirm that some readers did indeed use density to guide their decision-making~\citep{trumpbidenrecap}.

\section{Alternative approaches} \label{sec:alternative_approaches}

In developing the method described in the prior section, we also considered two alternative methods for diagnosing human error in image analysis, and describe them here---concluding that though they are intuitive they each suffer from drawbacks which render our primary approach preferable.

\subsection{Alternative approach 1: Train two models to predict ground truth and human judgment} \label{sec:twomodelsdifferenceocclusion} A straightforward algorithmic approach to diagnosing human error is to train one model to predict human judgment and a second model to predict ground truth, arguing that discrepancies between the two models indicate human error. To investigate this approach, we train one model to predict the aggregate human judgment $\hagg$, achieving an RMSE of 0.09 and a Spearman $r$ of 0.93; we train a second model to predict the ground truth continuous vote difference $\gtcont$, achieving an RMSE of 0.30 and a Spearman $r$ of 0.67 (Appendix \ref{sec:training_image_models}).\footnote{It is intuitive that prediction performance for $\hagg$ is better, because it is easier to predict: $\hagg$ should be almost entirely determined by the image, whereas it is unlikely that $\gtcont$ is.} We study how the two models differ by comparing the image patches which most change model predictions when they are masked out using the same occlusion mapping technique described in Section \ref{sec:whybias}. This method is direct and intuitive, and we show the results from it in Figure \ref{figure1}---revealing, for example, that the ground truth model associates road patches more strongly with Trump than does the human judgement model. In particular, two-lane highways divided by double-yellow lines are the most predictive of Trump neighborhoods, possibly because these highways signal the area is more rural.

However, this approach has several downsides. First, systematically comparing the image regions which influence two different deep learning models is difficult; identifying and interpreting salient features for even a \emph{single} model is a subtle and active area of research~\citep{adebayo2018sanity}. For example, if both models appear to be influenced by trucks, but the model predictions change by different amounts when trucks are occluded, it is unclear whether humans are misweighting trucks or the scales of the two model targets are simply incomparable. Second, this approach requires a ground truth model which outperforms human judgment; if humans perform better than the ground truth model, it is hard to argue that deviations from it are human mistakes. (Our main approach has a similar requirement: if humans outperform our estimated model of $p(\gt=1|\image)$, it is hard to argue that we are correctly identifying human error. However, we meet this requirement by design, by including the average human judgement as a feature in our model of $p(\gt=1|\image)$, and we verify that we outperform human judgment.) In this setting, we have enough data to train models that outperform human judgment, but in small data settings this may be difficult to do.

\subsection{Alternative approach 2: predict the difference between ground truth and human judgment}
\label{sec:single_model_predict_difference}
A second option is to train a model to directly predict the difference between ground truth and human judgment, $d_i = \hagg - \gt$, and then use occlusion mapping to identify image regions which contribute to a large difference. This method has several advantages over the method described in \Cref{sec:twomodelsdifferenceocclusion}: it is simpler, and it doesn't require a model which outperforms human judgment---just one which can predict the difference between ground and human judgment. However, it suffers from a conceptual flaw: this difference will be predictable even if humans are Bayes optimal---and so this method would incorrectly identify image features as causing errors even if humans are using them optimally. For example, suppose the only informative feature is whether the image has a car in it, and that 70\% of images with cars vote Trump while 20\% of images without cars vote Trump. Then, Bayes optimal humans will always classify images with cars as Trump (so $d = 0.3$ on average on car images) and images without cars as Biden (so $d = -0.2$ on average on non-car images). Our model will learn that cars predict the difference, implying that humans are over-weighting cars---even though in fact humans are Bayes optimal, with the error stemming from the fact that the images are not sufficiently informative. We view this conceptual flaw as sufficiently serious that we do not present results from this approach. We note that the method we favor in \Cref{sec:residual-interpretation}, which fits a model to estimate  $\residual = \hagg - m(\fhat)$ as opposed to $d_i = \hagg - \gt$, overcomes this limitation: if humans are Bayes optimal, $\residual$ will be uniformly 0 and we will not be able to identify image features which correlate with it. Observing this conceptual flaw was a primary motivation for  our method and in particular in developing a model to estimate $p(y_i | X_i)$.

We note that our favored method avoids the major weaknesses of both alternate approaches described above. To avoid having to compare occlusion maps between two different models, a weakness of the first alternate method described in \Cref{sec:twomodelsdifferenceocclusion}, our favored method examines the occlusion map from only a single model trained to predict the residual $\hagg - m(\fhat)$. To avoid the conceptual mistake of predicting the difference  $\hagg - \gt$, a weakness in the second alternate method described in \Cref{sec:single_model_predict_difference}, our favored method predicts the difference $\hagg - m(\fhat)$ rather than $\hagg - \gt$.

\section{Discussion} \label{sec:discussion}

In this work, we use a unique dataset of over 16 million human judgments with ground truth to propose a method for diagnosing human error in image analysis, a uniquely challenging setting for diagnosing human error. We show that by estimating $p(\gt=1|\image)$, we can decompose human error into bias, variance, and noise terms, and also identify specific image features which influence human judgment beyond the objective probability of the image $p(\gt=1|\image)$. We show that even if the machine learning model which forms a component of our estimate of $p(\gt=1|\image)$ is not perfectly optimal --- which will frequently be true for models trained on complex inputs like images especially in small-data regimes --- our approach can still provide useful insights into human error as long as the model adds signal beyond human judgment, a property we verify. We consider two alternate methods for diagnosing human error and assess their flaws. To facilitate reproduction and extension of our results, code to implement our method and reproduce our results is publicly available at \url{https://github.com/zamfi/diagnosing-human-error-in-image-analysis}.

\paragraph*{Limitations} There are several caveats in interpreting our results. First, we interpret our convolutional neural network models using standard and widely used interpretability techniques, but these are known to sometimes yield misleading conclusions~\citep{adebayo2018sanity,kindermans2019reliability}. Second, our data comes from self-selecting respondents to the \textit{New York Times} quiz, and as such the specific patterns we observe may not generalize to other populations. Although the methods we develop apply much more generally, any observed bias is only as representative as ``human bias'' as our sample of humans is representative of all humans. Third, our decomposition of human error into bias, variance and noise terms relies on our estimate of $p(\gt=1|\image)$ being approximately optimal, and we otherwise provide a \emph{lower bound} on the magnitude of human bias. While we provide suggestive evidence that our estimate of $p(\gt=1|\image)$ is reasonably close to optimal, we cannot verify this conclusively. Fourth, we cannot conclusively say that human error is due solely to the contents of the images humans are asked to evaluate; systematic errors can also be caused by poor task instructions or confusing user interface design, for example, and our method cannot isolate these effects specifically. Explicit variation in instructions or user interface design, if tracked, could serve as an additional variable to consider alongside or in conjunction with our method. 

\subsection*{Future directions}

There are many potential directions for future work.  Methodologically, there are several potential extensions to our current method. First, human decision-makers are heterogeneous: for example, previous work has developed methods for clustering humans by the errors they make~\cite{lakkaraju2016confusions}. It would be interesting to extend our method to model heterogeneity in human decision-makers. Second, we focus on the setting where ground truth labels are available for all observations, but in many real-world settings, ``selective labels'' mean that ground truth is censored by human judgments~\citep{kleinberg2018human,mullainathan2021diagnosing}: for example, we only observe a test result if a doctor decides to order a test. Extending our method to accommodate selective labels settings represents another avenue for future work.

Another direction for future work is applying our approach to other datasets. First, there are many other image datasets where our method could be applied: the ideal use case for our approach is an image dataset with human judgments $\hind$ and an objective ground truth label $\gt$ which is defined independently of human judgment. There is increasing recognition in the machine learning community that such objective ground truth labels (e.g., mortality in a medical setting) are invaluable to avoid merely laundering human biases into ``ground truth''~\citep{chen2020ethical}, and datasets are correspondingly becoming more widely available: for example, the Nightingale Open Science initiative\footnote{\url{https://docs.nightingalescience.org/}} is an effort to collect and make publicly available such datasets. Second, while we develop our method for image data, a more challenging setting than tabular data, many of our insights could equally be applied to tabular datasets---for example, the bias-variance-noise decomposition for human error we propose. Finally, our method is in principle applicable not just to human judgments, but to black-box algorithmic decision-making systems as well. We believe the method we propose is broadly applicable to diagnose and dissect both human and algorithmic error in a wide variety of settings.

\begin{acks}
The authors would like to thank \textit{The New York Times} for providing the initial dataset enabling this work and Michael Elabd for his contributions to an earlier version of this project. They thank Shengwu Li, the Cornell AI Policy and Practice group, and the anonymous reviewers for their thoughtful comments. Emma Pierson was supported by a Google Research Scholar Award and J.D. Zamfirescu-Pereira was partially supported by the United States Air Force and DARPA under contracts FA8750-20-C-0156, FA8750-20-C-0074, and FA8750-20-C0155 (SDCPS Program). These funders had no role in the design and conduct of the study; access and collection of data; analysis and interpretation of data; preparation, review, or approval of the manuscript; or the decision to submit the manuscript for publication.
\end{acks}

\bibliographystyle{ACM-Reference-Format}
\bibliography{sample-base}

\pagebreak

 \renewcommand\thefigure{\thesection\arabic{figure}}    
 \renewcommand\thetable{\thesection\arabic{table}}    
 \setcounter{figure}{0}  
 \setcounter{table}{0}   

\appendix

\section{Appendix}

\subsection{Data augmentation}
\label{sec:data_augmentation}
To better train our machine learning models, we augmented the \emph{New York Times} dataset with 52,025 additional images, alongside their ground truth vote shares.  

The official \emph{New York Times} dataset was constructed by randomly selecting 10,000 voter addresses for which precinct-level results were available, ensuring the sample was representative of the 2020 vote in both vote margin and population density~\citep{trumpbidengeography}. We replicate this methodology to acquire 52,025 additional images as follows, with the goal of producing an expanded dataset as similar as possible to the original \emph{New York Times} dataset (and indeed, we confirm as a robustness check that models trained on the original dataset yield similar performance on the expanded dataset).
\begin{enumerate}
    \item We start with a list of all electoral precincts, their geographic shapefiles, and their 2020 election results, compiled by \textit{The New York Times}: \url{https://github.com/TheUpshot/presidential-precinct-map-2020}. We draw a sample of precincts which matches \emph{The New York Times} sample on vote margin and population density because these were the variables used to rebalance \emph{The New York Times} sample.
    
    \item Using a proprietary voter file made available to us by an election analytics firm, for each precinct we sample up to 10 voters whose home address lies in the precinct.
    
    \item We use the Google Street View API to retrieve Street View images for each home location.
    
    \item Finally, to compensate for bias introduced by querying the Street View API (since not all home locations have Street View images), we again resample our dataset so it matches the original \emph{New York Times} data on vote margin and population density. (We confirm that the two datasets are also similar on other census demographics features like median age, household income, race/ethnicity, education, insurance levels, and home-ownership.)
\end{enumerate}

\begin{figure}[b]
    \centering
    \includegraphics[width=3in]{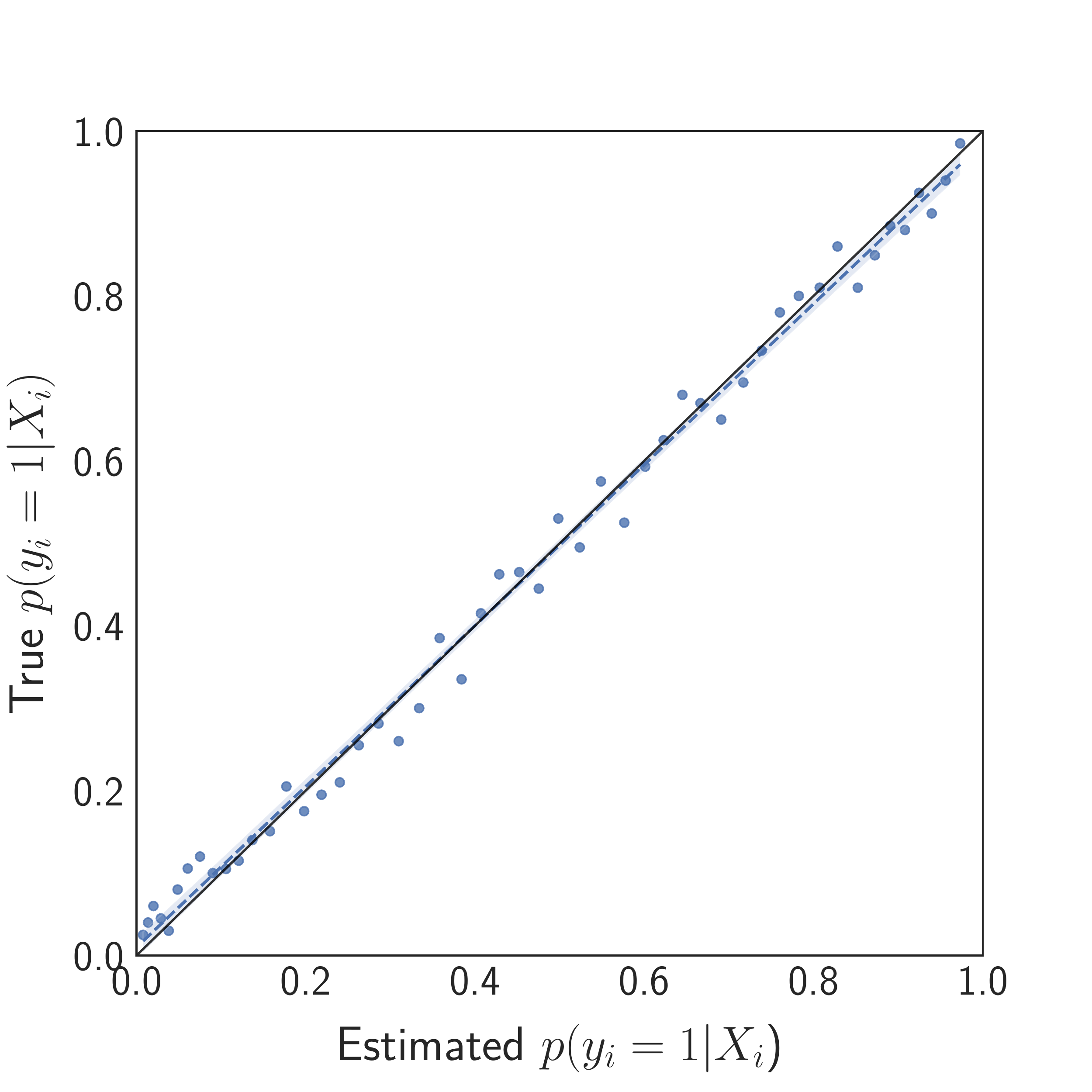}
    \caption{The model-estimated calibrated probabilities $\fhat = p(\gt=1|\image)$ (x-axis) line up well with the true probabilities, demonstrating that the model is calibrated. Observations are divided into 50 bins, sorting by $\fhat$; each point compares the mean values of $\fhat$ and $\gt$ in one bin.}
    \label{fig:calibration-plot}
\end{figure}

\begin{figure}[b]
    \centering
    \includegraphics[width=.9\linewidth]{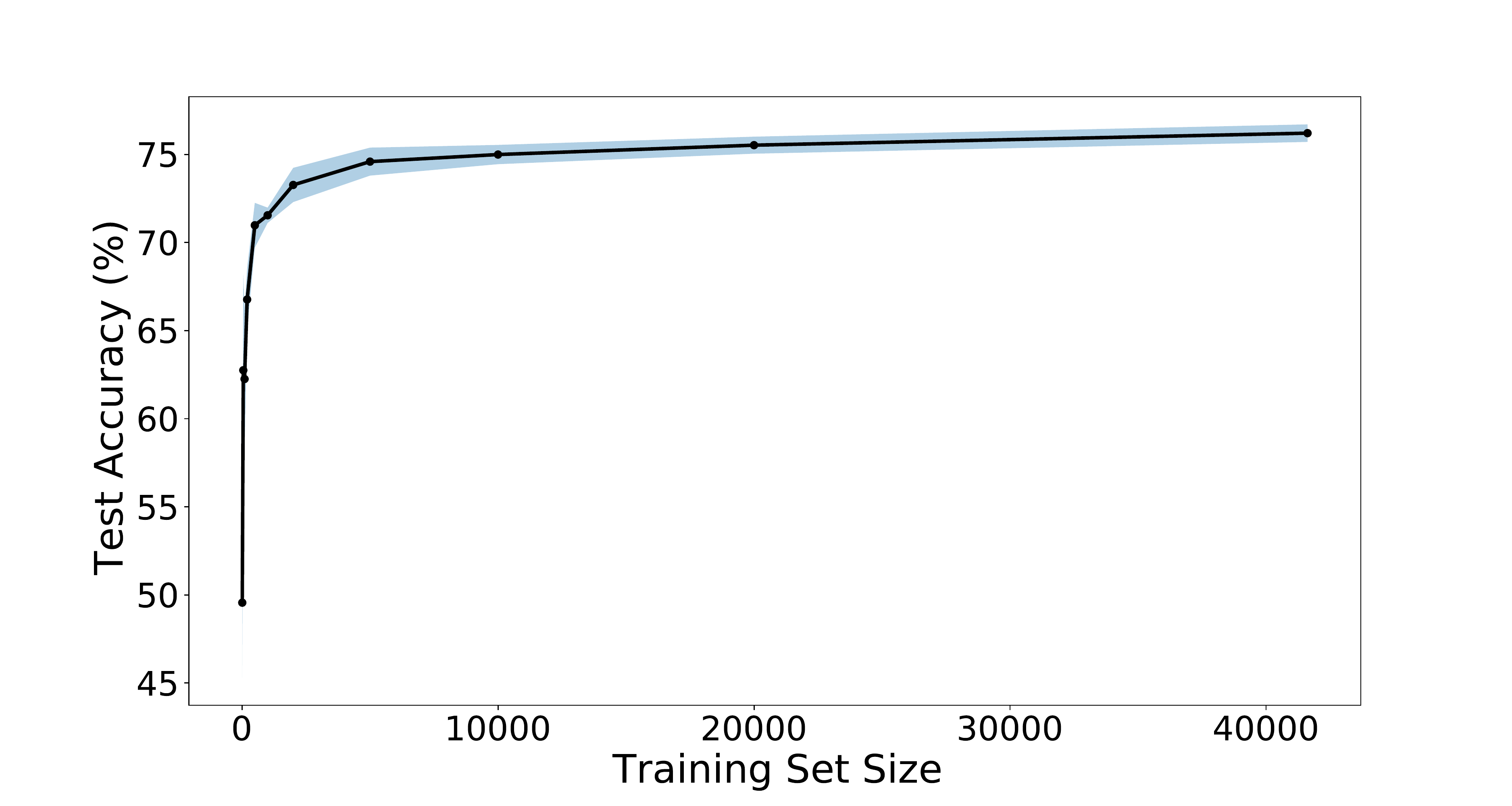}
    \caption{Performance of $\fhatext$ on training sets of different sizes. To reduce noise for small train sets, accuracy for each train set size is averaged across five randomly drawn train sets. Errorbars show the standard deviation across the five iterations. Model performance levels off as we approach the full train set size, suggesting that the train set size is large enough for model performance to approach optimality.}
    \label{fig:downsample_plot}
\end{figure}

\subsection{Training image models}
\label{sec:training_image_models}

We train deep learning models to predict three targets from Street View images: the aggregated human judgment $\hagg$, the continuous ground truth vote margin $\gtcont$, and the binary ground truth election outcome $\gt$.

\paragraph{$\hagg$ model.} To train a model to predict $\hagg$ from a neighborhood image, we begin with a ResNet model as the base model~\citep{he2015deep}. Because each neighborhood $X_i$ is represented by four images (corresponding to viewing angles of $0^\circ$, $90^\circ$, $180^\circ$, and $270^\circ$), our model architecture passes the images individually through the pre-trained ResNet and uses the final feature layer as the representation of each image. We then concatenate these representations to obtain a complete representation of all four images, which is then used as input into a series of fully-connected and ReLU layers to predict $\hagg$. We initialize the model with weights pre-trained on ImageNet~\citep{deng2009imagenet} and fine-tune the model on our dataset. We perform hyperparameter search over the ResNet architecture  (ResNet-34, ResNet-50, ResNet-101, or ResNet-152), the proportion of layers in the base ResNet model to unfreeze, optimizer parameters such as learning rate and decay, the resolution of the input images, and whether to randomly flip and crop the inputs. We train the $\hagg$ models on the \emph{New York Times} data with a mean squared error loss function and select the model with the lowest loss on the \emph{New York Times} validation dataset. Our model achieves 88.0\% accuracy on the holdout test set when $\hagg$ and the model predictions are binarized at 0.5, a Spearman $r$ of 0.93, and an RMSE of 0.09.

\paragraph{$\gtcont$ model.} Due to the similarity between the tasks of predicting $\hagg$ and $\gtcont$---both are predicting a continuous output from the neighborhood image as input---the model architecture and hyperparameter search remain the same. However, we train the $\gtcont$ models using the larger external training and validation sets before selecting the best model using the \emph{New York Times} validation set. (We cannot do this for the $\hagg$ model because we do not have data on $\hagg$ for the external dataset.) The highest-performing model that estimates $\gtcont$ achieves 74.6\% test accuracy when binarized at a threshold of 0, in other words predicting whether a neighborhood voted for Biden or Trump. In addition, the Spearman correlation between the predicted and true $\gtcont$ on the test set is 0.69, $p < 0.001$, and the RMSE is 0.29.

\begin{figure}[b]
    \centering
    \includegraphics[width=3in]{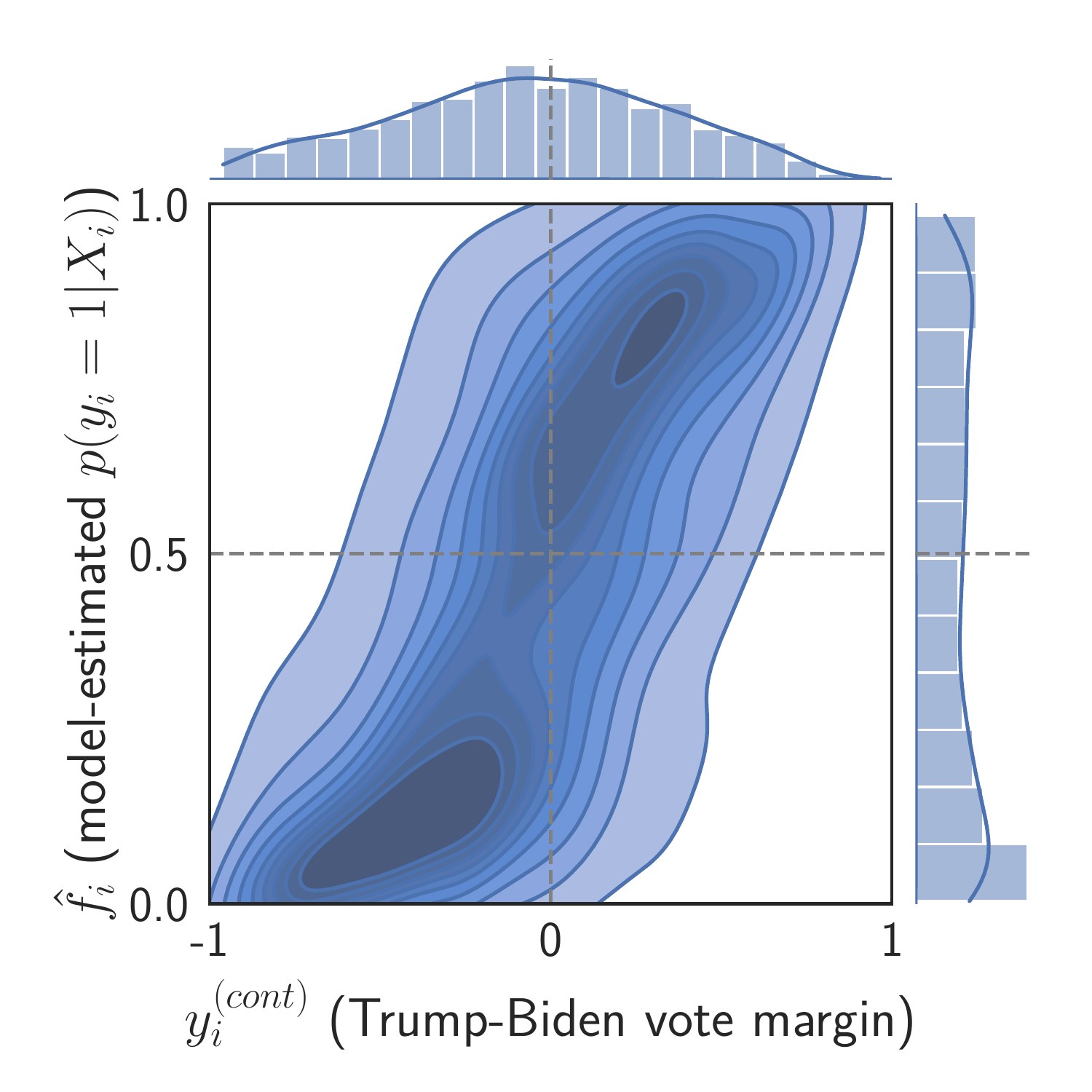}
    \caption{The model-estimated calibrated probabilities $\fhat = p(\gt=1|\image)$ (y-axis) are positively correlated, as expected, with the actual vote share $\gtcont$ (x-axis): the precincts where the vote was close are also those where the model expresses the greatest uncertainty.}
    \label{fig:gt-vs-yhat-calibrated}
\end{figure}

\paragraph{$p(\gt=1 | \image)$ model} The model, training, and dataset setup to estimate $p(\gt=1 | \image)$ are largely identical to that of the $\gtcont$ model. However, because $\gt$ is binary, we employ an additional sigmoid layer to convert the unbounded continuous output to a probability between 0 and 1. We also treat the task as a binary classification task to predict whether input neighborhoods voted for Biden (0) or Trump (1), rather than a continuous prediction task, and therefore use a negative log-likelihood loss function rather than MSE loss.
Our classifier achieves a test accuracy of 74.0\% and an AUC of 0.82.
A density plot comparing our modeled $p(\gt=1 | \image)$ with the ground truth $\gtcont$ appears in Figure~\ref{fig:gt-vs-yhat-calibrated}.

\paragraph{$r_i$ model.} Similarly, the model, training, and dataset setup to produce $\hat r_i$ are largely identical to that of the $\hagg$ model, except that the target for training is the residual $r_i = \hagg - m(\fhat)$ and we use the \emph{New York Times} training and validation datasets for model training and selection. We cannot use the external dataset to train the residual model because we do not have data on $\hagg$.

\begin{figure*}
  \centering
  \includegraphics[width=6in]{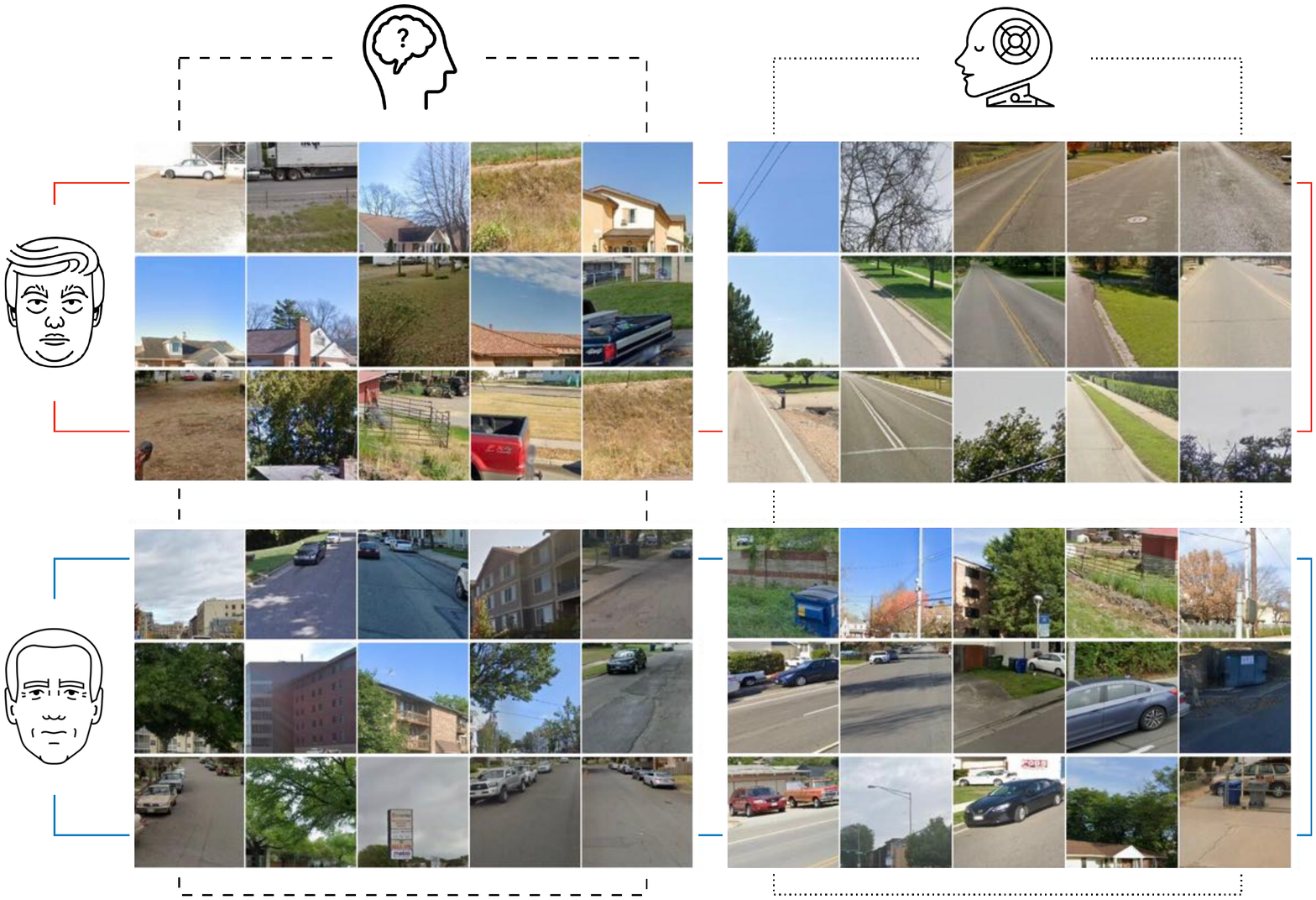}
  \caption{Image patches that most shift the model prediction towards Trump (top row) or Biden (bottom row), for the model which predicts the aggregate human judgment $\hagg$ (left column) and the model which predicts ground truth $\gtcont$ (right column). There is a clear visual difference between the left and right columns---for example, roads figure more prominently in the top right than the top left---indicating a difference in the features that most influence the ground truth model and the human judgment model. Underlying street view images \copyright{} Google.}
  \label{figure1}
\end{figure*}

\subsection{Identifying image regions most influencing prediction}
\label{sec:image_patches}

To identify image regions which influence a model's predictions (as in Figure \ref{figure1}) we mask out regions of the image and measure the change in model predictions, following previous work~\citep{zeiler2014visualizing}. Specifically, we divide each of the $0^\circ$, $90^\circ$, $180^\circ$, and $270^\circ$ Google Street View angles into a 4x4 grid, yielding a total of 64 square regions for each neighborhood image; for each square region, we measure how much the model prediction changes when we replace the square with a 60\% gray square. This yields a value for each square region which captures the impact of the region on the model's prediction. (We verify that the the results we report, e.g., finding ``open skies'', are robust to using a 2x2 grid size instead.)

In Figure \ref{figure1}, we show the regions which produce the largest changes in model outputs for the models predicting $\hagg$ and $\gtcont$, respectively. Comparing these regions corresponds to the method described in \Cref{sec:twomodelsdifferenceocclusion}.

\end{document}